\documentclass[prl,preprint,showpacs,showkeys,preprintnumbers,amsmath,amssymb]{revtex4-1}

\usepackage{amsmath,amssymb,bm,epsfig}
\usepackage{dcolumn}
\usepackage{bm}
\usepackage{feynmf}

\begin{document}

\preprint{Phys. Rev. B 96 235404 (2017)}

\title{Infrared Problem in Quantum Acoustodynamics\\
 at Finite Temperature}

\author{Dennis P. Clougherty}

\affiliation{
Department of Physics\\
University of Vermont\\
Burlington, Vermont 05405-0125}

\date{December 4, 2017}

\begin{abstract}
The phonon-assisted sticking rate of slow moving atoms impinging on an elastic membrane at nonzero temperature is studied analytically using a model with linear atom-phonon interactions, valid in the  weak coupling regime.  A perturbative expansion of adsorption rate in the atom-phonon coupling is infrared divergent at zero temperature, and this infrared problem is exacerbated by finite temperature. The use of a coherent state phonon basis in the calculation, however, yields infrared-finite results even at finite temperature.  The sticking probability with the emission of any finite number of phonons is explicitly seen to be exponentially small, and it vanishes as the membrane size grows, a result that was previously found at zero temperature; in contrast to the zero temperature case, this exponential suppression of the sticking probability persists even with the emission of an infinite number of soft phonons.  Explicit closed-form expressions are obtained for the effects of soft-phonon emission at finite temperature on the adsorption rate.  For slowly moving atoms, the model predicts that there is zero probability of sticking to a large elastic membrane at nonzero temperature and weak coupling.   


\end{abstract}

\maketitle
\section{Introduction}
The quantum adsorption of an atom to a vibrating, elastic membrane at zero temperature was studied in a previous paper \cite{dpc16}, where the phonon-assisted adsorption rate was calculated by treating the atom-phonon interaction as a perturbation.  The resulting rate was found to be well-defined and finite to lowest-order in the atom-phonon coupling; however,  if one carries on to higher orders, contributions were found to suffer from a divergence.   The origin of this divergence lies in contributions from the low-frequency phonons \cite{dpc16, dpc17}.      

It was shown that this infrared problem can be remedied \cite{dpc16} by evaluating the transition matrix elements with a coherent state (CS) basis for the phonons.  This basis is rather natural because if one includes the interaction of the adsorbed atom with the phonons in the unperturbed Hamiltonian $H_0$, the CS basis is an eigenbasis for $H_0$.    

The use of the CS phonon basis yields transition matrix elements for adsorption containing a phonon overlap factor that reduces the matrix element.  As a result,  adsorption to the membrane is quenched for processes involving the creation of any finite number of phonons; however, the total adsorption rate, which includes the creation of an infinite number of soft phonons, was found to be finite and nonzero, a situation familiar from bremsstrahlung scattering in QED \cite{landaulifshitz4} where only the emission of an infinite number of photons gives a nonzero scattering rate.

Here, the theory is generalized to adsorption to a membrane that initially has a thermal distribution of phonons.  As a result of the enhanced presence of low-energy phonons in a membrane at nonzero temperature, adsorption on a membrane has an infrared problem that is more severe than at zero temperature: the loop expansion of the atom self-energy with the atom-phonon interaction has infrared divergences at every order.

Despite the increase in severity, this infrared problem at finite temperature is also completely cured with the use of the CS phonon basis.  It will be shown explicitly that the adsorption rate for processes creating a finite number of phonons in a membrane at nonzero temperature is also quenched.  There is, however,  a major difference in the adsorption at finite temperature:  even with the inclusion of adsorption by the emission of an infinite number of soft phonons, the adsorption rate remains quenched.  Thus, a sufficiently slowly moving neutral atom will not stick to a large elastic membrane, even though this process is energetically allowed.  It will be shown that this lack of adsorption is a result of a phonon orthogonality catastrophe that suppresses the adsorption transition. This finding supports a result that was obtained previously by variational means \cite{dpc13}.  

\section{Model}
The dynamics of a slow atom impinging upon an elastic membrane at normal incidence is described by a Hamiltonian \cite{dpc13} of the form $H=H_a+H_m+H_i$ with
\begin{eqnarray}
H_a&=&{P^2\over 2 M}+V(Z)\nonumber\\
H_m&=&\sum_{n=1}^N \omega_{n}b^\dagger_{n}b_{n}\nonumber\\
H_i&=& \sum_{n=1}^N {\cal U}(Z) (b_{n}+b^\dagger_{n})
\label{eq:model}
\end{eqnarray}
where $M$ is the mass of the impinging atom, $Z$ is the distance of the atom above the equilibrium plane of the membrane, and $V(Z)$ is the interaction of the atom with a static membrane. $b_n$   $(b_n^{\dagger})$ annihilates (creates) a  phonon in the  membrane with energy $\omega_n$. The atom-phonon interaction ${\cal U}$ can be expressed in terms of physical parameters of the membrane \cite{dpc13}.

The effects of the corrugation of the potential $V$ due to the discrete translational symmetry of a 2D crystal can be trivially included in this model.  The soft-phonon effects, the focus of this work, would be unchanged by corrugation, since the soft-phonon effects studied here depend only on having a nonvanishing atom-phonon coupling to the adsorbate.

The Hilbert space for the atom is partitioned into disjoint subspaces with a projection operator \cite{feshbach2} $\hat{\cal P}$ that projects out only the bound state $|B\rangle$ of $H_a$.  $\hat{\cal Q}$ is the complementary projection operator such that $\hat{\cal P} +\hat{\cal Q}=1$.  $H$ is then separated into $H_0+H_1$ where

\begin{eqnarray}
H_0&=&{P^2\over 2 M}+V(Z)+\sum_{n} \omega_{n}b^\dagger_{n}b_{n}+\hat{\cal P} H_i \hat{\cal P}\\
H_1&=& \hat{\cal Q} H_i \hat{\cal P}+ \hat{\cal P} H_i \hat{\cal Q}+\hat{\cal Q} H_i \hat{\cal Q}
\end{eqnarray}
$H_0$ can be uncoupled using a unitary transformation $U=e^{-S}$ where
\begin{equation}
S=\sum_n {g_{bb}\over\omega_n}(b_{n}-b^\dagger_{ n}) \hat{\cal P} 
\end{equation}
and $g_{bb}=\langle B|{\cal U}|B\rangle$.

The resulting transformed Hamiltonian $\tilde H_0=U^\dagger H_0 U$ is
\begin{equation}
\tilde H_0={P^2\over 2 M}+V(Z)+\sum_{n} \omega_{n}b^\dagger_{n}b_{n}-\hat{\cal P}\Delta 
\label{eq:ibm}
\end{equation}
where $\Delta=\sum_n {g_{bb}^2\over\omega_n}$.

The model of Eq.~\ref{eq:model} might be termed quantum acoustodynamics (QAD), as it describes  the coupling of low-energy atoms to phonons. The infrared behavior of QAD has similarities to the infamous infrared problem in QED \cite{Bloch:1937pw,IR-coherent,Yennie:1961ad}; in QED, the infrared divergences have their origin in the low-frequency photons.

The eigenstates of $\tilde{H}_0$ are a direct product of an atomic eigenstate and a phonon Fock state.  The ground state of $\tilde{H}_0$ is then $|B\rangle\otimes|\{0\}\rangle$ where $|B\rangle$ is the lowest atomic eigenstate of ${P^2\over 2 M}+V(Z)$ (taken to be a bound state with energy $-E_b$) and $|\{0\}\rangle$ is the phonon vacuum of Eq.~\ref{eq:ibm}.  

The corresponding eigenstate of ${H}_0$ is obtained with the application of $U$.
Thus, the ground state of $H_0$ ($U|B;\{0\}\rangle$)  has energy $-E_b-\Delta$, with the shift $\Delta$ due to the phonon interaction with the adsorbed atom.  The eigenstates of $H_0$ with an adsorbed atom are phonon coherent states.  A one-phonon eigenstate of $\tilde H_0$ with an adsorbed atom $|B;1_q\rangle$ generates a ``phonon-added'' CS as the corresponding eigenstate of $H_0$ with energy $-E_b-\Delta+\omega_q$.  

\section{Sticking rate}

The adsorption process can be regarded as a capture reaction \cite{dpc92, 06ecapture} of the form $A\to A_b+\phi$ where $A_b$ is the bound atom and $\phi$ is a phonon created in the membrane.  Taking an initial phonon distribution of $\{n_q\}$ in the membrane with the unbound atom, the  reaction rate under the perturbation $H_1$ is then written as 
\begin{eqnarray}
\Gamma_1&=2\pi \sum_{s, \{n_q\}}   |\langle {k};\{n_q\}|H_1 U|B;\{n_q\}_{q\ne s}, n_s+1\rangle|^2 \cr
&\times p\big(\{n_q\}\big)\delta\big(\Omega_0-\omega_s\big)
\end{eqnarray}
where $\Omega_0=E+E_b+\Delta$ (with $E=k^2/2M$) and $p\big(\{n_q\}\big)$ is the probability of a phonon occupancy distribution of $\{n_q\}$ at temperature $T$.   
There is an implicit constraint on the maximum atom energy allowed, since the energy transfer to the membrane $\Omega_0$ must be less than the Debye frequency $\omega_D$.  Thus, 
$E< \omega_D-E_b-\Delta$.  For atomic hydrogen adsorbing on a micromembrane of graphene, the parameters have been assumed to have the following approximate values \cite{dpc17}:  $\omega_D\approx  65$ meV, $E_b\approx 40$ meV, and $\Delta\approx 10$ meV. 
This constrains the physical parameter regime to incident energies below $E< 15$ meV.  

Details on the evaluation of the  transition matrix element are given in the Appendix.  The following expression results
\begin{equation}
\Gamma_1=2 \pi g_{kb}^2(k)\rho\ {\cal R}(E_s, \epsilon)\ Q(\Omega_0, T) P_N(\epsilon, T)
\label{gam1}
\end{equation}
$\rho$ is the density of membrane vibrational modes, $E_s=E+E_b$, and $g_{kb}=\langle {k}|{\cal U}|B\rangle$, a model parameter that depends on  the initial wavevector  $k$ of the atom.  $\epsilon$, the lowest phonon frequency supported by the membrane, serves as a practical infrared regulator for the model.  It has the added benefit of describing finite-sized membranes.

The other dimensionless factors ${\cal R}$, $Q$ and $P_N$ are defined and described in what follows.  As a check for $\Gamma_1$ in Eq.~\ref{gam1}, the zero temperature rate obtained in Ref.~\cite{dpc16} is immediately recovered, since  $Q\to 1$ and $P_N\to \exp{(-\sum_p \lambda_p^2)}$ as $T\to 0$ (see Appendix).

The reduction factor ${\cal R}(E_s, \epsilon)$ appeared previously in the zero temperature calculation and was defined as
\begin{equation}
{\cal R}(E_s, \epsilon)=\bigg(1+{\Delta(\epsilon)\over E_s}\bigg)^{-2}
\end{equation}

\begin{figure}[h!]
\includegraphics[width=.8\columnwidth]{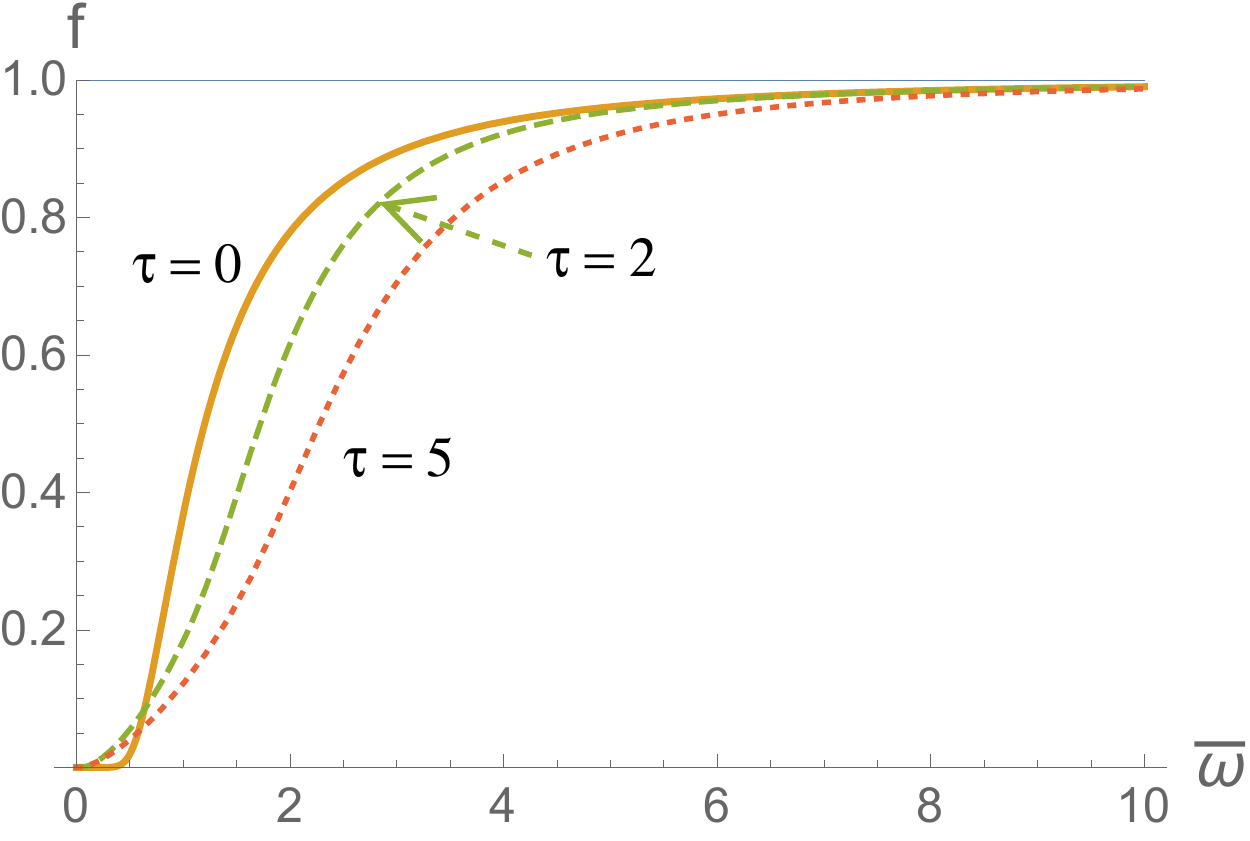} 
\caption{\label{fig:overlap} The phonon matrix element overlap $f$ vs. the dimensionless frequency $\bar\omega$ for $\tau=0$ (solid),  2 (dashed),  5 (dotted). The matrix element overlap $f$ is (1) bounded ($0\le f < 1$) for a phonon spectrum $0\le\omega\le\omega_D$ and (2) monotonically increasing in $\omega$.}
\end{figure}

\begin{figure}[h!]
\includegraphics[width=.8\columnwidth]{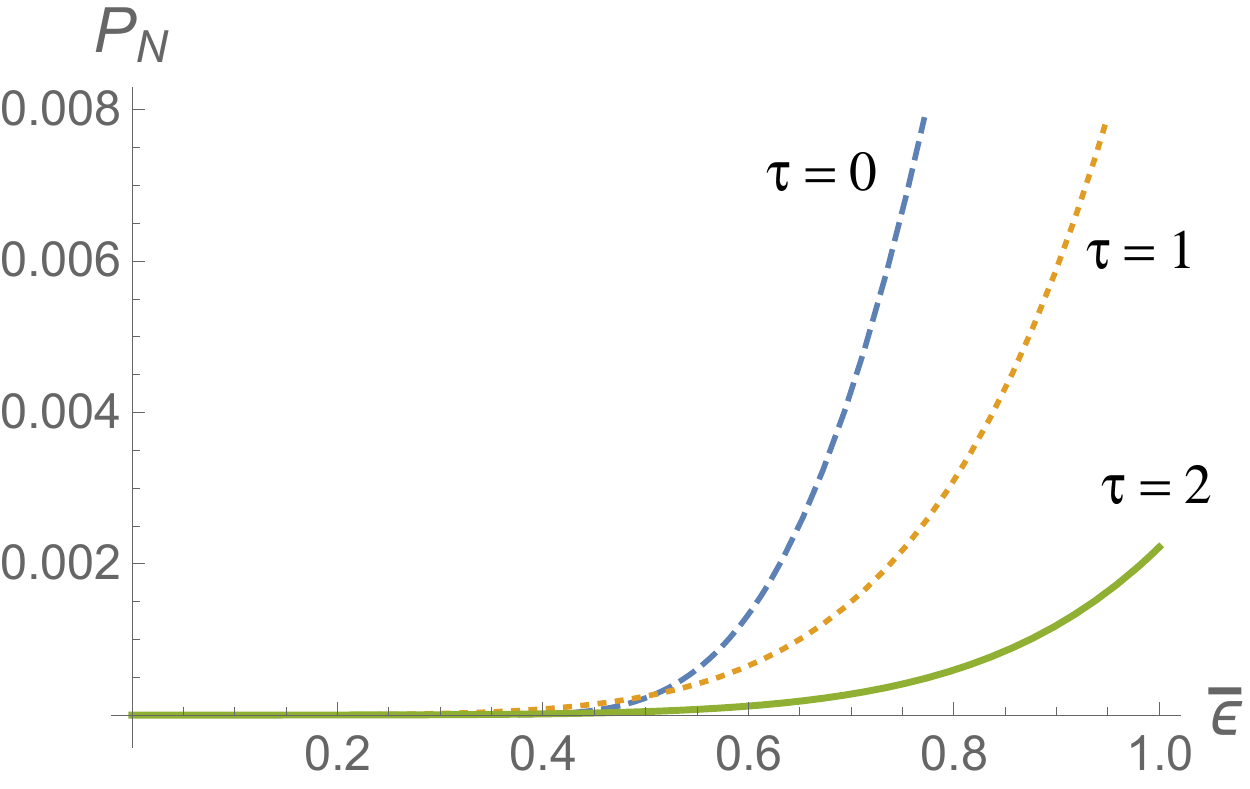} 
\caption{\label{fig:PNeps} The phonon reduction factor $P_N$ vs. the dimensionless IR cutoff $\bar\epsilon$ at  $\tau=0$ (dashed), 1 (dotted), 2 (solid) for $N=30$ vibrational modes.  }
\end{figure}

\begin{figure}[h!]
\includegraphics[width=.8\columnwidth]{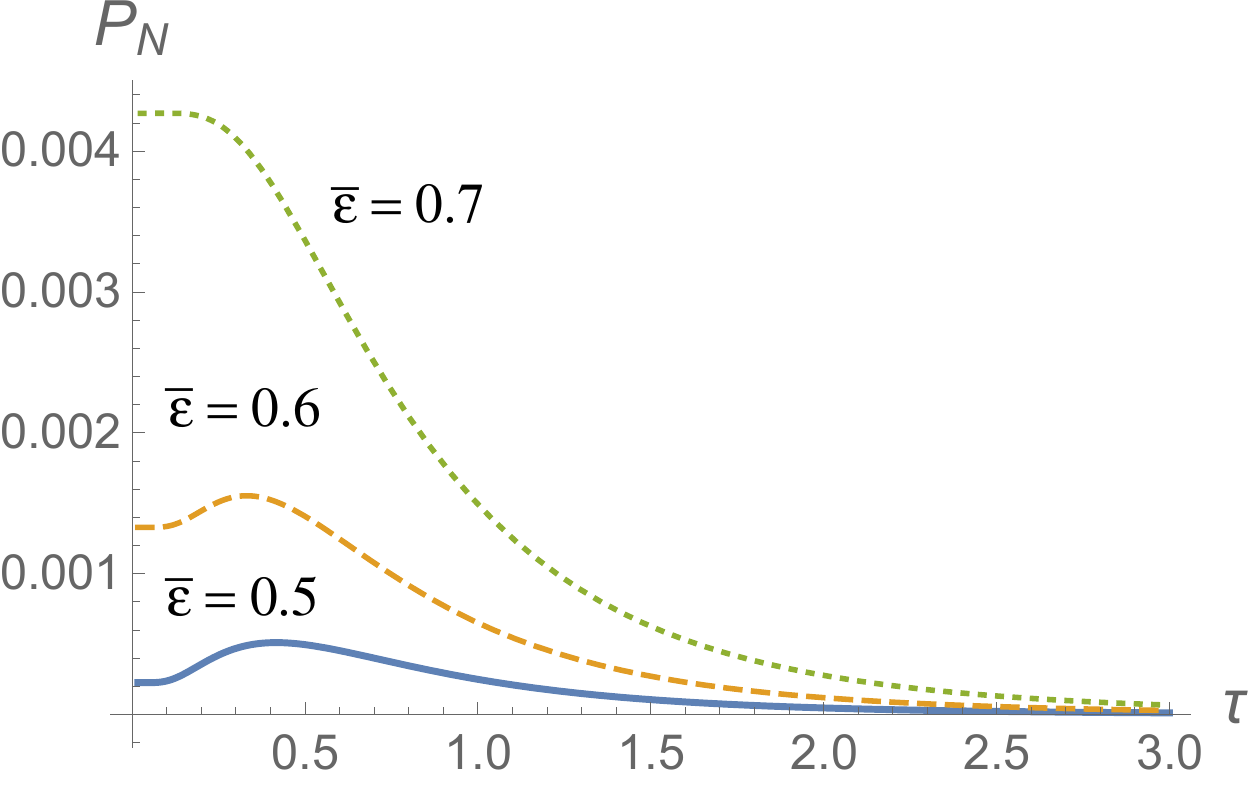} 
\caption{\label{fig:PNtau} The phonon reduction factor $P_N$ vs dimensionless membrane temperature $\tau$ at  $\bar\epsilon=0.5$ (solid), 0.6 (dashed), 0.7 (dotted) for $N=30$ vibrational modes. Local maxima at nonvanishing $\tau$ are visible for $\bar\epsilon =$ 0.5 and 0.6.}
\end{figure}

\begin{figure}[h!]
\includegraphics[width=.8\columnwidth]{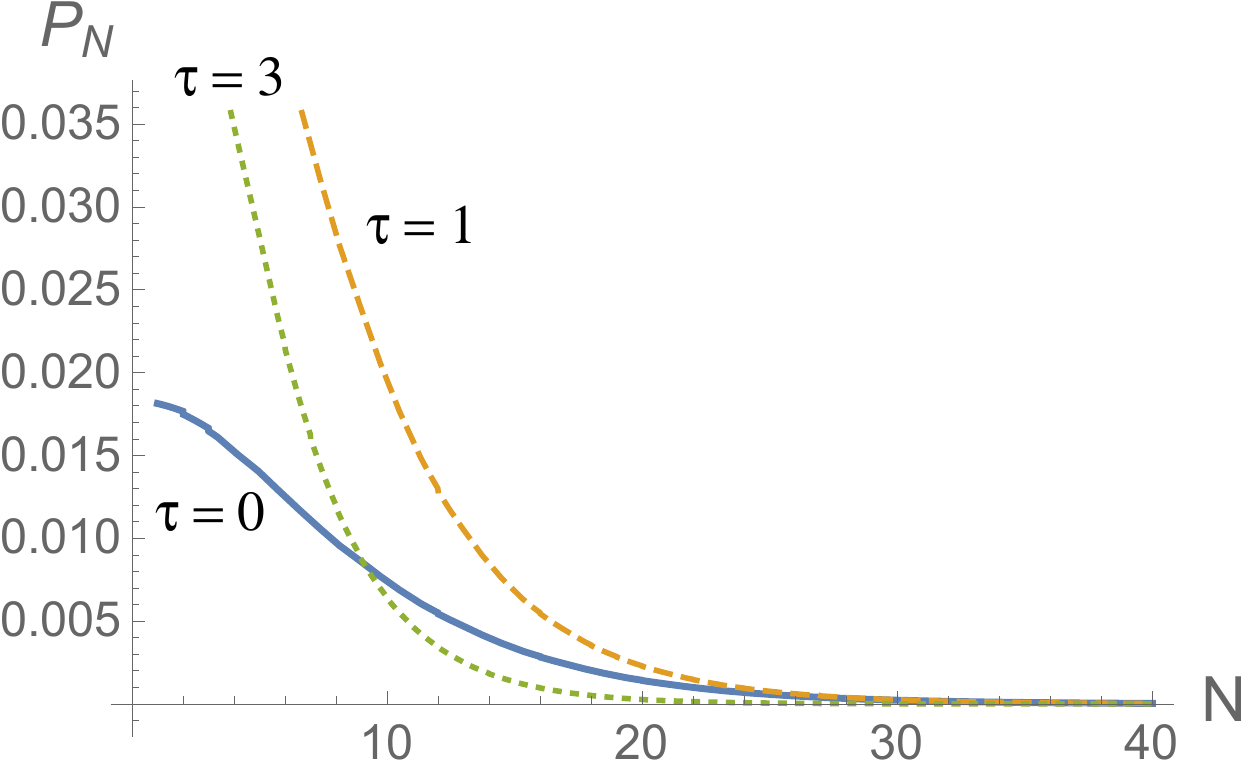} 
\caption{\label{fig:PNN} The phonon reduction factor $P_N$ vs the number of vibrational modes $N$ at  $\bar\epsilon=0.5$ for $\tau=0$ (solid), 1 (dashed), and 3 (dotted).  }
\end{figure}

$P_N(\epsilon, T)$ is the phonon reduction factor given by
\begin{equation}
P_N(\epsilon, T)\equiv\prod_{p=1}^N f(\bar\omega_p, \tau)=\prod_{p=1}^N e^{- \bar\omega_p^{-2}(2 \bar{n}_p+1)} I_0\big(2(\bar{n}_p+1)\bar\omega_p^{-2} e^{-{\bar\omega_p\over 2\tau}}\big)
\label{eq:PN}
\end{equation}
where $\tau=T/g_{bb}$ and ${\bar\omega_p}=\omega_p/g_{bb}$.  $I_0$ is the modified Bessel function of the first kind of order zero.  The phonon overlap factor $f$ for a single mode as function of frequency $\omega$ is plotted in Fig.~\ref{fig:overlap}.  The behavior of the phonon reduction factor $P_N$ for the membrane is given in Figs.~\ref{fig:PNeps}--\ref{fig:PNN}.  $\Gamma_1$ vanishes exponentially in the IR limit of $\epsilon\to 0$ as a result of the phonon reduction factor $P_N$ (see Fig.~\ref{fig:PNeps}). 

The factor $Q$ is defined as
\begin{equation}
Q(\omega, T)\equiv {\bar\omega^2 e^{{\bar\omega\over 2\tau}}}{I_1(2(\bar{n}+1)\lambda^2 e^{-{\bar\omega\over 2\tau}})\over I_0(2(\bar{n}+1)\lambda^2 e^{-{\bar\omega\over 2\tau}})}
\label{eq:Q}
\end{equation}
Here, $\lambda= \bar\omega^{-1}=g_{bb}/\omega$, $\tau=T/g_{bb}$, and $\bar{n}=(\exp({\beta \omega})-1)^{-1}$.

\begin{figure}[h!]
\includegraphics[width=.8\columnwidth]{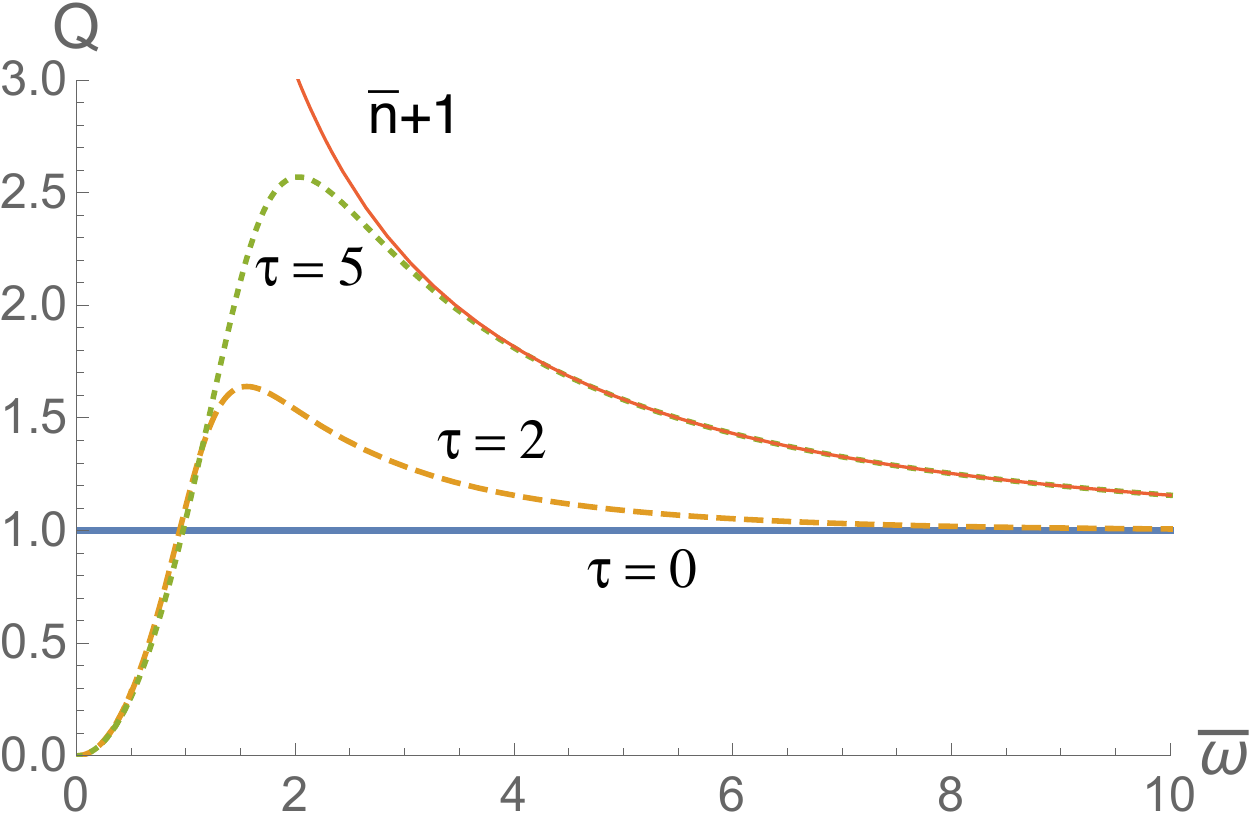} 
\caption{\label{fig:Q}  $Q$ vs $\bar\omega$ at $\tau=0$ (solid), 2 (dashed), and 5 (dotted).  For large $\bar\omega$, $Q\sim(\bar{n}(\bar\omega,\tau)+1)$ (thin red).  At nonzero temperatures,  $Q\sim\bar\omega^2$ for $\bar\omega\to 0$. }
\end{figure}

$Q(\omega, T)$ is a factor containing the enhancement for sticking by stimulated phonon emission due to the presence of thermal phonons.  In Fig.~\ref{fig:Q}, one sees that $Q$ behaves as $\bar{n}+1$ for large $\omega$.  For low frequencies, however, $Q$ suppresses the sticking rate relative to the zero temperature case. 

The behavior of $Q$ with frequency is shown in Fig.~\ref{fig:Q}.  For binding energies that exceed the temperature, $Q$ approaches $\bar{n}+1$, the expected factor from both stimulated and spontaneous phonon emission.  For small binding energies, however, $Q$ varies as $\omega^2$ which suppresses the adsorption rate.  The reduction of $Q$ at low frequencies is a result of a reduced  phonon emission rate for an adsorbate that is dressed by a cloud of soft phonons.   
The phonons acquire a small thermal mass through the interaction with the adsorbate that suppresses $Q$ at low frequencies and cuts off singular contributions to adsorption from soft-phonon emission.

For a final state with two phonons emitted, the two-phonon sticking rate $\Gamma_2$  to ${\cal O}(g_{kb}^2)$ is
\begin{equation}
\Gamma_2=2 \pi\sum_{m < n} \sum_{\{n_q\}}p({\{n_q\}}) |\langle B;\{n_q\}_{q\ne n, m}, n_n+1, n_m+1|e^S H_1|{ k};{\{n_q\}}\rangle|^2 \delta(\Omega_0-\omega_m-\omega_n) 
\end{equation}
This yields 
\begin{eqnarray}
\Gamma_2&=&2\pi g_{kb}^2 P_N(\epsilon, T) \sum_{p < q} {(\lambda_p+\lambda_q-\lambda_p\lambda_q\Lambda)^2} 
Q(\omega_p,T)Q(\omega_q, T) \delta(\Omega_0-\omega_p-\omega_q)\nonumber\\
&\approx&\Gamma_1 \int_\epsilon^{\omega_s} {d\omega \over\omega^2} \rho g_{bb}^2 Q(\omega, T)
\label{eq:gamma2}
\end{eqnarray}
where $\Lambda=\sum_k \lambda_k$ and $\omega_s$ ($\ll \Omega_0$) serves as the upper limit for the soft-phonon regime.  As a check of Eq.~\ref{eq:gamma2}, it is noted that the zero temperature result for $\Gamma_2$ [Eq.~[A12] in Ref.~\cite{dpc16}] follows immediately since $Q(\omega, T=0)=1$.  

Since $Q(\omega, T)\sim \omega^2$ as $\omega\to 0$ for nonzero $T$, the integral is finite as $\epsilon\to 0$, signaling that the use of a CS basis for phonons remedies the infrared problem for $T\ne 0$.  

There is a temperature-dependent frequency scale that separates the regimes of  $Q$.  The low-frequency regime has $2(\bar{n}(\omega, T)+1)\bar{\omega}^{-2} e^{-{\omega\over 2T}}\gg 1$.  Thus  $\omega_c$, defined by $2(\bar{n}(\omega_c, T)+1)\bar{\omega}_c^{-2} e^{-{\omega_c\over 2T}}\sim 1$, can be used to delineate the regimes of $Q$.  For $g_{bb}\ll T \ll E_b$, $\omega_c\approx(2 g_{bb}^2 T)^{\frac{1}{3}}$. 
For the case of atomic hydrogen impinging on a suspended micromembrane of graphene, this would only constrain the membrane temperature $T$ to values less than room temperature. 

For $\bar\omega_c\gg 1$, Eq.~\ref{eq:gamma2} gives
\begin{equation}
\Gamma_2\approx \Gamma_1 \rho \omega_c
\end{equation}

For adsorption by emission of $n$ phonons, a hard phonon ($\omega\sim \Omega_0$) and $n-1$ soft phonons ($\omega\sim\omega_c$), the adsorption rate $\Gamma_n$ is given by
\begin{equation}
\Gamma_n\approx  \Gamma_1 {(\rho \omega_c)^{n-1}\over (n-1)!}
\end{equation}
This generalization is consistent with treating soft-phonon radiation as a Poisson process where the emissions are independent events.

The total rate resulting from a final state with an arbitrary number of soft phonons is obtained by summing over the partial rates.  Thus,
\begin{equation}
\Gamma=\sum_{n=1}^\infty \Gamma_n \approx \Gamma_1 e^{\rho \omega_c}
\end{equation}
Consequently, the total adsorption rate  is infrared-finite. For small, nonzero $\epsilon$, the adsorption rate is exponentially small, and it vanishes in the limit of $\epsilon\to 0$.

\section{Discussion}

It is worthwhile to compare the case of a membrane initially at $T=0$  with one at nonzero T.  While the loop expansion at $T\ne 0$ contains infrared divergences of greater severity than at $T=0$, the use of a CS phonon basis in the evaluation of transition matrix elements yields IR-finite results in both cases.  

In both cases, there is an exponential reduction  in the adsorption rate for $\epsilon\to 0$ in processes with the emission of a finite number of phonons; however, the total adsorption rate, obtained by summing over all possible soft phonon emissions,  vanishes only logarithmically as the IR cutoff $\epsilon\to 0$.  In contrast, in the case of $T\ne 0$, the total adsorption rate remains exponentially suppressed as $\epsilon\to 0$. This exponential suppression of the total adsorption rate is consistent with previous results using a variational mean-field method that show the suppression of adsorption for $T\ne 0$ for sufficiently weak atom-phonon coupling strength \cite{dpc13}. 


Finally, the system studied here provides a mechanical realization for a recent reinterpretation of the infrared problem in QED where ``non-trivial scattering''  \cite{strominger17} results in vacuum transitions that render matrix elements IR-finite.  In QAD, there are an infinity of  degenerate phonon vacua connected by the unitary transformation $U$ that uniformly displaces the membrane.  The initial phonon vacuum is orthogonal to the phonon vacuum for an atom bound to the membrane, a phonon orthogonality catastrophe \cite{dpc11}. 

Thus far, it appears that the emphasis of numerical quantum chemical studies of atom-surface sticking and scattering from 2D materials has been on the calculation of the atomic transition matrix elements $g_{kb}$, and the consequences of the vacuum transition have not been included.  There are several recent numerical studies that omit the effect of the phonon vacuum transition in quantum sticking to graphene \cite{lepetit-jackson,bonfanti2}.  As a result, these numerical studies obtain nonvanishing sticking probabilities at low incident energies.

\vfil\eject

\appendix*
\section{Appendix}
\subsection{The transition matrix element for adsorption}

The transition matrix element between the initial state and the bound state is  
\begin{eqnarray}
{\cal M}_p\big(\{n_q\}\big)&\equiv&\langle B;\{n_q\}_{q\ne p},n_p+1|e^S H_1|{ k};\{n_q\}\rangle\nonumber\\
&=&\sum_n g_{kb} \langle \{n_q\}_{q\ne p},n_p+1|X^\dagger (a_n+a_n^\dagger)|\{n_q\}\rangle
\label{eq:M}
\end{eqnarray}
where $X^\dagger=\exp\big(-\sum_q \lambda_q (a_q^\dagger-a_q)\big)$ and $g_{kb}({k})=\langle {k}|{\cal U}|B\rangle$.

For convenience, we define 
\begin{equation}
Z(n', n)\equiv \langle n'|{\cal T}^\dagger (a+a^\dagger)|n\rangle
\end{equation}
and
\begin{equation}
T(n', n)\equiv \langle n'|{\cal T}^\dagger |n\rangle
\end{equation}
where ${\cal T}^\dagger=\exp\big(-\lambda (a^\dagger-a)\big)$.
Thus,

\begin{eqnarray}
\langle \{n_q\}_{q\ne p},n_p+1|X^\dagger (a_k+a_k^\dagger)|\{n_q\}\rangle= \delta_{p k} Z(n_p+1,n_p)\prod_{q\ne p} T(n_q, n_q)\cr
 + (1-\delta_{p k}) Z(n_k, n_k) T(n_p+1, n_p) \prod_{q\ne p, k} T(n_q, n_q)
\end{eqnarray}

One can express these quantities in a simple form by using commutation relations for the phonons and recursion relations  \cite{g&r} for generalized Laguerre polynomials $L^{(\alpha)}_n$.  To summarize, 
\begin{equation}
T(n_q, n_q)=e^{-\lambda_q^2/2} L_{n_q}(\lambda_q^2)
\end{equation}
\begin{equation}
T(n_q+1, n_q)=-e^{-\lambda_q^2/2} {\lambda_q\over\sqrt{n_q+1}}L^{(1)}_{n_q}(\lambda_q^2)
\end{equation}

\begin{equation}
Z(n_p+1, n_p)=
\begin{cases}
e^{-\lambda_p^2/2}{(1-\lambda_p^2)\over\sqrt{n_p+1}}L^{(1)}_{n_p}(\lambda_p^2), & n_p\ge 1\\
e^{-\lambda_p^2/2}L_{1}(\lambda_p^2), & n_p = 0
\end{cases}
\end{equation}
\begin{equation}
Z(n_p, n_p)=\lambda_p e^{-\lambda_p^2/2}L_{n_p}(\lambda_p^2)
\end{equation}

By combining these identities, one straightforwardly obtains the following expression for the transition matrix element
\begin{equation}
{\cal M}_p\big(\{n_q\}\big)=
{g_{kb}{\cal P}\big(\{n_q\}\big) \over\sqrt{n_p+1}}{L^{(1)}_{n_p}(\lambda^2_p)\over L_{n_p}(\lambda^2_p)} \big(1-\lambda_p\sum_m \lambda_m\big)
\end{equation}
where 
\begin{equation}
{\cal P}\big(\{n_q\}\big)\equiv \prod_q e^{-\lambda_q^2/2} L_{n_q}(\lambda_q^2)
\end{equation}

The zero temperature result [Eq.~(A7) in Ref.~\cite{dpc16}] follows by setting the initial phonon occupancies $\{n_q\}$ to zero.  The limit of vanishing $g_{bb}$ also provides a check on the expression for ${\cal M}_p$: taking all $\lambda\to 0$ yields
\begin{eqnarray}
{\cal M}_p\big(\{n_q\}\big)&=&
{g_{kb} \over\sqrt{n_p+1}}{(n_p+1)!\over n_p!}\nonumber\\
&=&g_{kb} \sqrt{n_p+1}
\end{eqnarray}
since $L^{(\alpha)}_n(0)={n+\alpha \choose n}$.  The same result follows immediately by setting $X^\dagger\to 1$ in Eq.~\ref{eq:M}.

 The transition rate $\Gamma_1$ in Eq.~(\ref{gam1}) is obtained using the following identities \cite{g&r} to evaluate thermal averages
\begin{equation}
\langle L^2_{n_p}(\lambda_p^2) \rangle_T=\exp{\big(-2\bar{n}_p\lambda_p^2\big)} I_0\big[2(\bar{n}_p+1)\lambda_p^2\exp{\big(-{\beta\omega_p\over 2}\big)}\big]
\end{equation}

\begin{eqnarray}
\bigg\langle \bigg({L^{(1)}_{n_p}(\lambda_p^2)\over \sqrt{n_p+1}}\bigg)^2 \bigg\rangle_T&={1\over\lambda_p^2}\exp{\big({\beta\omega_p\over 2}\big)}\exp{(-2\bar{n}_p\lambda_p^2)}\cr
&\times I_1\big[2(\bar{n}_p+1)\lambda_p^2\exp{\big(-{\beta\omega_p\over 2}\big)}\big]
\end{eqnarray}
where $\bar{n}_p=(\exp({\beta\omega_p})-1)^{-1}$.

For two-phonon emission,
the transition matrix element ${\cal M}_{p\ell}$ $(p\ne\ell)$ is  
\begin{eqnarray}
{\cal M}_{p\ell}\big(\{n_q\}\big)&\equiv&\langle B;\{n_q\}_{q\ne p, \ell},n_p+1, n_\ell+1|e^S H_1|{ k};\{n_q\}\rangle\nonumber\\
&=&{g_{kb}({k}){\cal P}\big(\{n_q\}\big) \over\sqrt{(n_p+1)(n_\ell+1)}}{L^{(1)}_{n_p}(\lambda^2_p)\over L_{n_p}(\lambda^2_p)} {L^{(1)}_{n_\ell}(\lambda^2_\ell)\over L_{n_\ell}(\lambda^2_\ell)}\big(\lambda_\ell+\lambda_p-\lambda_p\lambda_\ell\sum_k \lambda_k\big)
\end{eqnarray}

\vfil\eject

\subsection{Some properties of the phonon reduction factor}

The phonon reduction factor $P_N(\epsilon, T)$ is defined as
\begin{equation}
P_N(\epsilon, T)\equiv \prod_{p=1}^N f(\bar\omega_p, \tau)=\prod_{p=1}^N e^{- (2 \bar{n}_p+1)/\bar\omega_p^2} I_0(2(\bar{n}_p+1)\bar\omega_p^{-2} e^{-{\bar{\omega}_p\over 2\tau}})
\end{equation}
where $p$ runs over the $N$ axisymmetric vibrational modes of the membrane.

As $I_0(x)\sim 1$ as $x\to 0$, the zero temperature limit of Eq.~\ref{eq:PN} is given by
\begin{equation}
\lim_{T\to 0} P_N(\epsilon, T)=\exp{(-\sum_{p=1}^N \lambda_p^2)}
\end{equation}
This is the factor $\exp{(-2F)}$ obtained in Eq.~(7) of Ref.~\cite{dpc16}.  Also, since $I_1(x)\sim x/2$ as $x\to 0$,  Eq.~\ref{eq:Q} gives  $\lim_{T\to 0} Q(\omega, T)=1$.

In the continuum approximation, 
\begin{eqnarray}
P(\epsilon, T=0)&\approx&\exp{\bigg(-\int_\epsilon^{\omega_D}d\omega {\rho g_{bb}^2\over\omega^2}\bigg)}\cr
&\approx&\exp{\bigg(-{\rho g_{bb}^2\over\epsilon}\bigg)}
\end{eqnarray}


An upper bound to the  phonon reduction factor at nonzero temperature is obtained from the  asymptotic behavior of $I_0$ for large argument in Eq.~\ref{eq:PN}.  Thus,
\begin{eqnarray}
P_N(\epsilon, T)&<&f(\bar\epsilon, \tau)\cr
&\sim&\exp{\bigg(-{1\over 4\bar\epsilon\tau}\bigg)}\sqrt{\bar\epsilon^3\over 4\pi\tau},\ \ \ \epsilon\to 0
\end{eqnarray} 
This rapid decrease in $P_N$ with decreasing $\bar\epsilon$ is seen in Fig.~\ref{fig:PNeps}.  


\bibliographystyle{apsrev4-1}
\bibliography{qst}

\end{document}